\documentclass[aps,prb, reprint, superscriptaddress,preprintnumbers,amssymb]{revtex4-2}
\usepackage{graphicx}% Include figure files
\usepackage{dcolumn}% Align table columns on decimal point
\usepackage{bm}% bold math

\begin{document}

\preprint{ver5.1}

\title{Transport evidence for twin-boundary pinning of superconducting vortices in FeSe}


%\author{}
%\thanks{}
%\affiliation{}

\author{Taichi Terashima}
\email{TERASHIMA.Taichi@nims.go.jp}
\affiliation{Research Center for Materials Nanoarchitectonics (MANA), National Institute for Materials Science, Tsukuba 305-0003, Japan}
\author{Hideaki Fujii}
\affiliation{Department of Physics, Okayama University, Okayama 700-8530, Japan}
\author{Yoshitaka Matsushita}
\affiliation{Research Network and Facility Services Division, National Institute for Materials Science, Tsukuba 305-0047, Japan}
\author{Shinya Uji}
\affiliation{Research Center for Materials Nanoarchitectonics (MANA), National Institute for Materials Science, Tsukuba 305-0003, Japan}
\author{Yuji Matsuda}
\affiliation{Department of Physics, Kyoto University, Kyoto 606-8502, Japan}
\author{Takasada Shibauchi}
\affiliation{Department of Advanced Materials Science, University of Tokyo, Kashiwa, Chiba 277-8561, Japan}
\author{Shigeru Kasahara}
\email{kasa@okayama-u.ac.jp}
\affiliation{Research Institute for Interdisciplinary Science, Okayama University, Okayama 700-8530, Japan}


\date{\today}

\begin{abstract}
We provide bulk transport evidence for twin-boundary pinning of vortices in FeSe.
We measure interlayer resistance in FeSe in magnetic fields and find that, as the field is rotated in the $ab$ plane, the flux-flow resistivity is suppressed when the field direction is parallel to twinning planes.
The width of the associated dip in the resistance vs in-plane field direction curve varies as $T^{1/2}B^{-3/4}$, consistent with the creation of kinked vortices near the parallel field geometry.
\end{abstract}

%\keywords{}

\maketitle

%\section{}
%\subsection{}
%\subsubsection{}

% If in two-column mode, this environment will change to single-column
% format so that long equations can be displayed. Use
%\begin{widetext}
%\end{widetext}

% figures should be put into the text as floats.
% Use the graphics or graphicx packages (distributed with LaTeX2e)
% and the \includegraphics macro defined in those packages.
% See the LaTeX Graphics Companion by Michel Goosens, Sebastian Rahtz,
% entire page. There is no need to do explicit centering.

\section{Introduction}
Twin boundaries in superconductors are known to affect behavior of superconducting vortices by suppressing or enhancing superconductivity near them.
A famous example is flux pinning by twin boundaries in YBa$_2$Cu$_3$O$_{7-\delta}$ (YBCO) single crystals \cite{Kwok90PRL, Roitburd90PRL, Gyorgy90APL}.
It was reported that the flux-flow resistivity was largely suppressed when the magnetic field was parallel to the twinning planes.

Most iron-based superconductors, unless sufficiently doped, undergo a tetragonal-to-orthorhombic structural phase transition as cooled from room temperature.
Accordingly, crystals are twinned below the transition temperature $T_s$.
The twin boundaries are parallel to high-temperature tetragonal (100) or (010) planes.
Scanning SQUID and magnetic force microscopy studies on Ba(Fe$_{1-x}$Co$_x$)$_2$As$_2$ and BaFe$_2$(As$_{1-x}$P$_x$)$_2$ have reported that twin boundaries enhance superfluid density and repel vortices \cite{Kalisky20PRB, Yagil16PRB, [{It was also reported that the critical current density was enhanced when the density of twin boundaries was large. }]Prozorov09PRB}.
By contrast, scanning tunneling microscopy studies on FeSe have shown that twin boundaries in FeSe suppress the superconducting gap and superfluid density, and pin vortices \cite{Song12PRL, Watashige15PRX}.
The twin-boundary pinning of vortices in FeSe has also been confirmed in a scanning SQUID study \cite{Zhang19PRB}.
Fe(Se, Te) is a prime candidate in a search for Majorana fermions \cite{Wang18Science, Machida19NatMater}, and these twin boundaries may be used to arrange vortices as desired in quantum computing applications where vortices carrying Majorana fermions are manipulated \cite{Song23NanoLett}.
Therefore twin-boundary pinning properties in FeSe may be of interest.
In this work, we provide bulk transport evidence for twin-boundary pinning of vortices in FeSe and show that temperature and magnetic-field variation of pinning properties can be described by a theory previously developed for YBCO \cite{Blatter91PRB}.

\begin{figure}
\includegraphics[width=8.6cm]{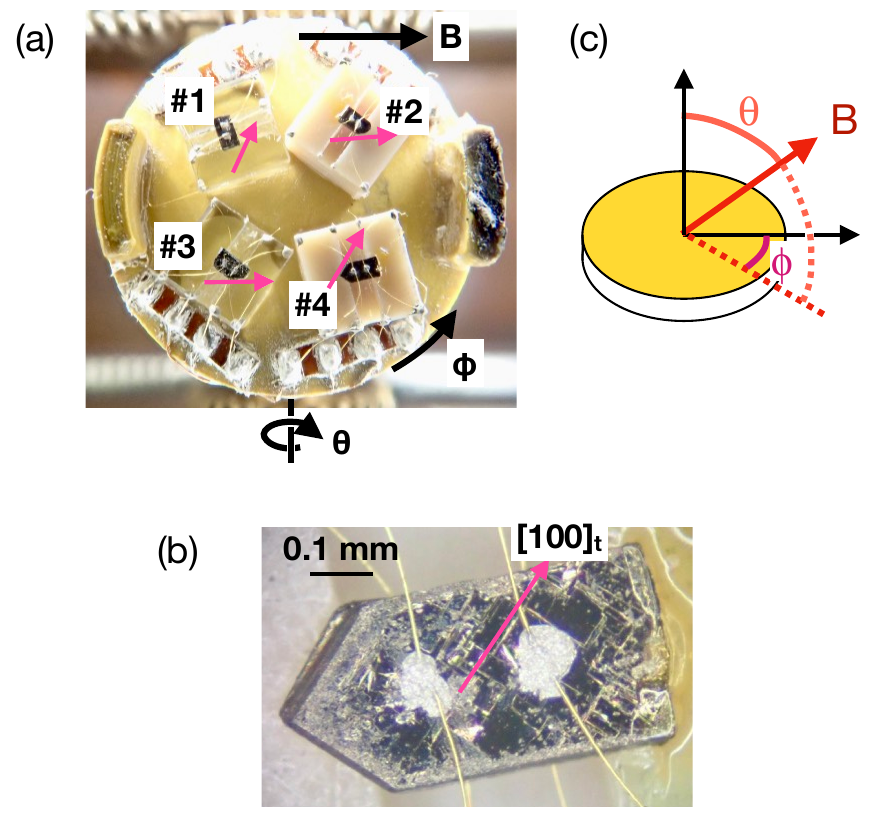}
\caption{\label{Sample}Sample platofrom and samples.
(a) Two-axis rotation platform at $\theta = 90^{\circ}$ and $\phi = 0^{\circ}$.
The diameter of the platform is 12.4 mm.
Four samples \#1--4 are mounted, for each of which the [100]$_t$ direction is indicated by pink arrows.
(b) Blow-up of sample \#4.
The [100]$_t$ direction is readily determined from the morphology of the cleaved surface.
(c) The polar $\theta$ and azimuthal $\phi$ angles of the applied field were defined with respect to the platform.
}
\end{figure}

\section{Experiments}
High-quality single crystals of FeSe were grown by a chemical vapor transport method \cite{Bohmer13PRB}.
Figure 1 shows the experimental setup:
Four samples were mounted on a rotation platform of the two-axis rotator probe.
To measure interlayer resistance $R_c$, a current and a voltage contact were spot-welded on each (001) plane and then reinforced by silver conducting paste.
Notice that the [100]$_t$ direction, where the subscript $t$ refers to the room-temperature tetragonal cell, is easily recognized from the surface morphology [Fig. 1(b)].
The twin boundaries run along the [100]$_t$ and [010]$_t$ directions.
The polar $\theta$ and azimuthal $\phi$ angles of the applied magnetic field were defined with respect to the platform as shown in Fig. 1(c).
To remove possible Hall voltage contamination, measurements were performed at a positive field $+B$ and at a negative one $-B$ for each field direction, and the symmetrized voltage was used to calculate $R_c$, i.e. $R_c = (V(+B) + V(-B) )/2I$, although the antisymmetric voltage $(V(+B) - V(-B) )$ was on average less than 1\% of the symmetric one.
In the following, we concentrate on samples \#3 and 4, on which high-quality data were obtained (for samples \#1 and 2, see Appendix A).
For these samples, the orientation of the crystal axes were confirmed by X-ray diffraction measurements after all the resistance measurements were finished.

\begin{figure}
\includegraphics[width=8.6cm]{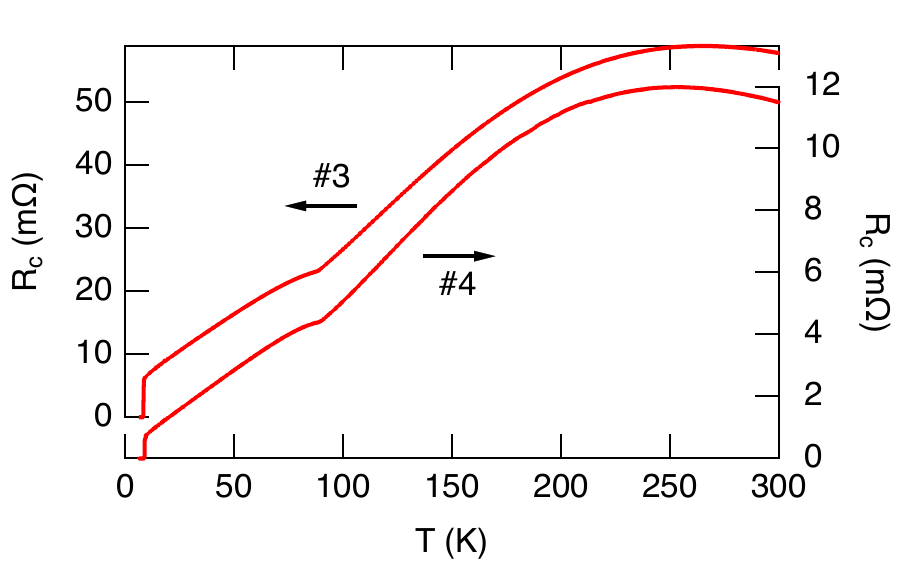}
\caption{\label{RvsT}
Interlayer resistance $R_c$ vs temperature for samples \#3 and 4.
}
\end{figure}

\begin{figure}
\includegraphics[width=8.6cm]{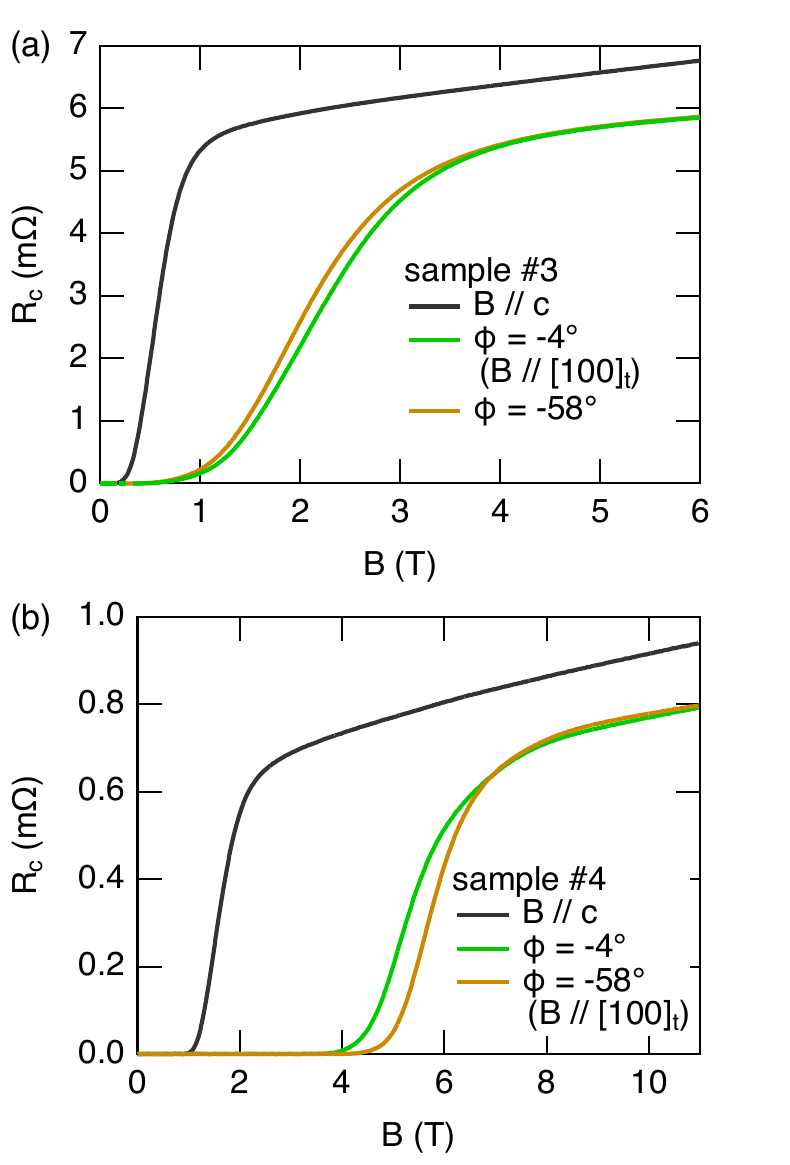}
\caption{\label{RvsB} Magnetoresistance of samples \#3 and 4 at $T$ = 8 K for three field directions, $B \parallel c$ and $B \bot c$ with $\phi$ = -4 and -58$^{\circ}$.
$\phi$ = -4$^{\circ}$  corresponds to $B \parallel [100]_t$ for sample \#3 (a), while $\phi$ = -58$^{\circ}$ does for sample \#4 (b).  Notice different horizontal scales for (a) and (b), which is the main reason for the apparent broad transitions in sample \#3 (a).}
\end{figure}

\section{Results and discussion}
Figure 2 shows the temperature dependence of the interlayer resistance $R_c$ for samples \#3 and 4.
While the temperature dependence of the in-plane resistance is metallic from room temperature \cite{Kasahara14PNAS}, the measured interlayer resistance curves exhibit a non-metallic temperature dependence, i.e., d$R_c$/d$T < 0$, near room temperature.
The structural transition temperature $T_s$, superconducting transition temperature $T_c$, and residual resistivity ratio at $T$ = 10 K are 87.6 K, 8.5 K, and 8.7 for sample \#3, and 89.4 K, 9.0 K, and 15 for sample \#4, respectively.
The interlayer resistivity at $T$ = 10 K is estimated to be 0.50, 1.8, and 0.57 m$\Omega$~cm for sample \#2, 3, and 4, respectively, which is roughly comparable to a value of $\sim$1 m$\Omega$~cm reported in \cite{Kaluarachchi16PRB}.
For comparison, the in-plane resistivity at $T$ = 10 K is roughly 15 $\mu \Omega$~cm \cite{Kasahara14PNAS, Terashima16PRB2}.
Therefore, the resistivity anisotropy is estimated to be more than 30.

Figure 3 shows the field dependence of the interlayer resistance $R_c$ at $T$ = 8 K for samples \#3 and 4.
The black curves are for $B \parallel c$, whereas the green and brown ones are for in-plane fields.
For sample \#3, the resistance in the superconducting transition region is smaller at $\phi = -4^{\circ}$ (green) compared to $\phi = -58^{\circ}$ (brown), while for sample \#4 it is smaller at $\phi = -58^{\circ}$ (brown).
Looking at the pink arrows in Fig. 1(a), we notice that $\phi = -4^{\circ}$ and $-58^{\circ}$ correspond to $B \parallel [100]_t$ for samples \#3 and 4, respectively, as far as the naked eye can see.

\begin{figure}
\includegraphics[width=8.6cm]{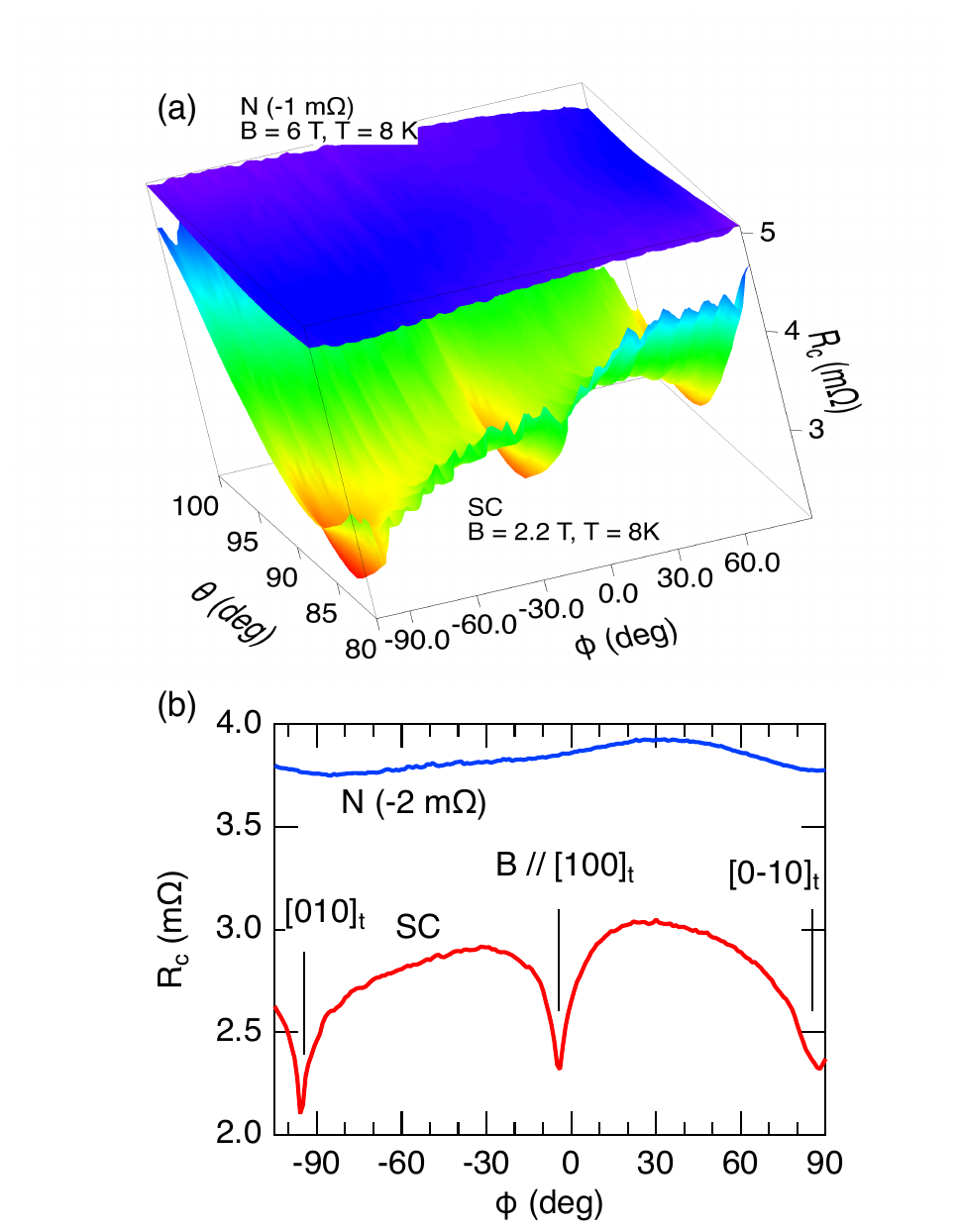}
\caption{\label{ch3} Interlayer resistance $R_c$ of sample \#3 in magnetic fields.
(a) Resistance as a function of $\phi$ and $\theta$ measured at $B$ = 2.2 T and $T$ = 8 K in the superconducting transition region (SC) compared to that at $B$ = 6 T and $T$ = 8 K in the normal state (N).
The latter is offset by -1 m$\Omega$.
(b) Resistance for in-plane fields as a function of $\phi$ for the superconducting transition region (SC) and normal state (N).
The normal state curve is offset by -2 m$\Omega$.}
\end{figure}

\begin{figure}
\includegraphics[width=8.6cm]{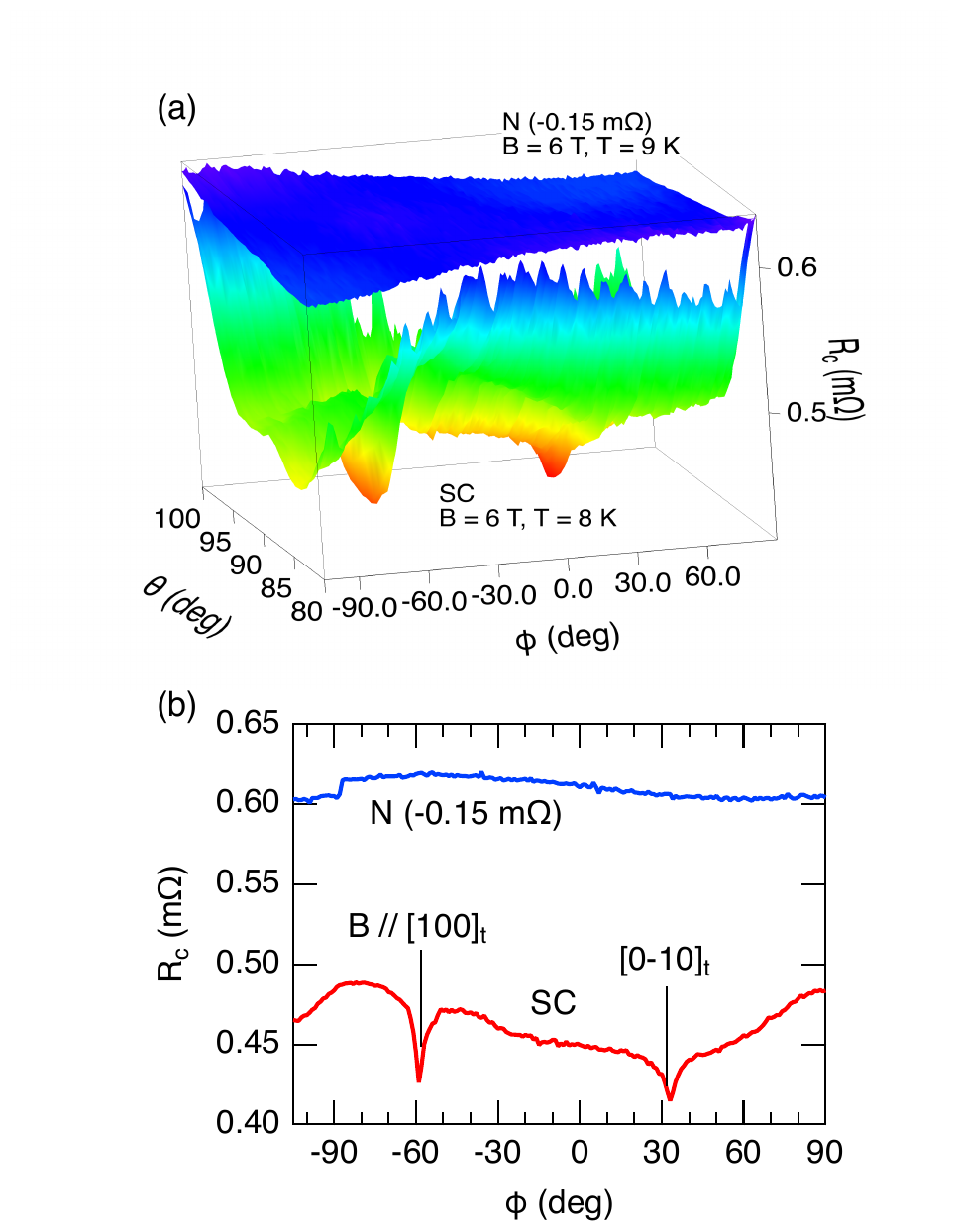}
\caption{\label{ch4} Interlayer resistance $R_c$ of sample \#4 in magnetic fields.
(a) Resistance as a function of $\phi$ and $\theta$ measured at $B$ = 6 T and $T$ = 8 K in the superconducting transition region (SC) compared to that at $B$ = 6 T and $T$ = 9 K in the normal state (N).
The latter is offset by -0.15 m$\Omega$.
(b) Resistance for in-plane fields as a function of $\phi$ for the superconducting transition region (SC) and normal state (N).
The normal state curve is offset by -0.15 m$\Omega$.}
\end{figure}

Figures 4 and 5 show detailed field-orientation dependence of the interlayer resistance in the superconducting transition region and in the normal state.
Figure 4(a) shows the interlayer resistance of sample \#3 measured in the transition region at $B$ = 2.2 T and $T$ = 8 K as a function of $\theta$ and $\phi$ compared to that in the normal state at $B$ = 6 T and $T$ = 8 K.
The current density is 0.46 A/cm$^2$.
We also performed measurements with the current density of 0.15 A/cm$^2$ but observed no appreciable change.
The transition-region data show three distinct dips.
Considering a slight misalignment of the sample, we picked up the minimum resistance at each $\phi$ as the resistance for the in-plane field and plotted it as a function of $\phi$ in Fig. 4(b).
The dip positions correspond to the field directions parallel to [010]$_t$, [100]$_t$, and [0$\bar{1}$0]$_t$ within experimental accuracy.
The background variation of the resistance outside the dip regions may be ascribed to inhomogeneous current distribution, i.e., the current is not exactly along the $c$ axis everywhere in the sample.
We also note that, to allow two-axis rotation, electrical wires are not fixed and hence that small pickup of electromotive force induced by wire vibration due to ac current is inevitable.

Figure 5(a) shows the interlayer resistance of sample \#4 measured in the transition region at $B$ = 6 T and $T$ = 8 K as a function of $\theta$ and $\phi$ compared to that in the normal state at $B$ = 6 T and $T$ = 9 K.
The current density is 0.51 A/cm$^2$.
The resistance for the in-plane field is shown as a function of $\phi$ in Fig. 5(b).
The dip positions correspond to the field directions parallel to [100]$_t$, and [0$\bar{1}$0]$_t$ within experimental accuracy.

\begin{figure}
\includegraphics[width=8.6cm]{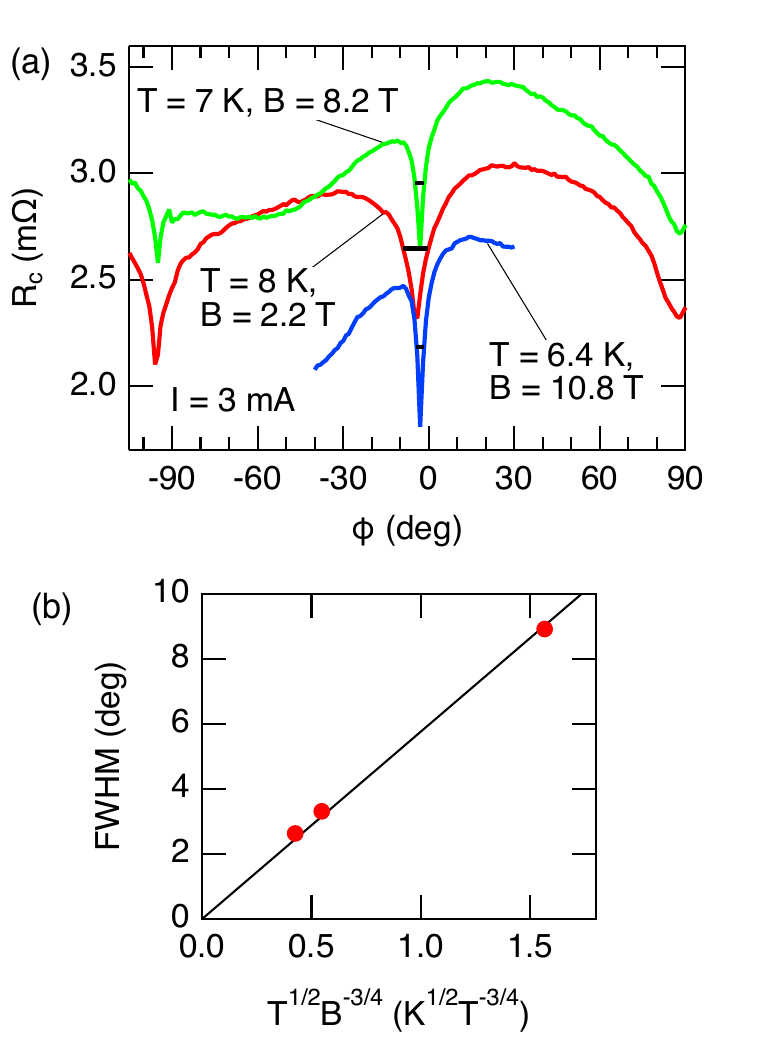}
\caption{\label{ch3_3T} 
Temperature and field dependence of the resistance dip.
(a) Interlayer resistance $R_c$ of sample \#3 for in-plane fields as a function of $\phi$ for three different temperatures and fields as indicated.
The black horizontal lines indicate FWHM for the $\phi \sim 0$ dips.
(b) FWHM plotted against $T^{1/2}B^{-3/4}$.
The solid line indicates the linear relation between the two quantities.
}
\end{figure}

Figure 6(a) shows the resistance of sample \#3 for in-plane field as a function of $\phi$ measured at different temperatures and hence different field strengths.
The dip sharpens as the temperature is lowered and hence the field is increased.
We determined the full width at half maximum (FWHM) of the $\phi \sim 0$ dip for each of the three curves in Fig. 6(a) (for details, see Appendix B) and plotted it against $T^{1/2}B^{-3/4}$ in Fig. 6(b).
We see a nice linear relation between the two quantities.

Our observations can qualitatively be explained as follows:
When $B \parallel [100]_t$ (or [010]$_t$), some vortices are trapped by twin boundaries.
Because $I \parallel c$, the direction of the Lorentz force is $[010]_t$ ([100]$_t$), i.e., perpendicular to the boundaries.
However, as long as the twin-boundary pinning force is stronger than the Lorentz force, the vortices do not move and hence do not contribute flow resistance.
Because the vortex spacing $l$ is much smaller than the twin-boundary spacing $d$ as explained below, only a portion of vortices is trapped by the twin boundaries, and hence we observe a resistance drop, not zero resistance.

The inequality $l \ll d$ is justified as follows:
For a triangular vortex lattice in isotropic superconductors, the vortex spacing is given by $a_{\Delta}=1.074(\Phi_0/B)^{1/2}$, where $\Phi_0$ is the flux quantum.
Considering the coherence length anisotropy of $\xi_{ab}/\xi_{c} \sim 4$ \cite{Terashima14PRB}, the in-plane vortex spacing $l$ is estimated to be $l = \sqrt{\xi_{ab}/\xi_{c}}a_{\Delta} \sim$ 66--30 nm for $B$ = 2.2--10.8 T.
On the other hand, photoemission electron microscopy (PEEM) \cite{Shimojima21Science} and STM \cite{Song23NanoLett} observations of twin boundaries in FeSe suggest that a typical twin-boundary spacing can be assumed to be $d \sim$ 300 nm.

As already noted at the beginning of this article, vortex pinning due to twin boundaries was previously studied in the high-transition-temperature cuprate superconductor YBCO\cite{Kwok90PRL, Roitburd90PRL, Gyorgy90APL}.
The present data are reminiscent of \cite{Li93PRB}, where the interlayer resistance of YBCO in the superconducting transition region was measured as a function of in-plane field direction and resistance drops were observed for field directions parallel to the twin boundaries.

Blatter \textit{et al.} gave a first theoretical description of the twin-boundary pinning in YBCO \cite{Blatter91PRB}: 
The authors argued that the twin boundaries attract the vortices due to the suppressed order parameter and hence that each vortex deforms to adjust itself to the twinning planes.
Accordingly, as long as the angle $\Delta\phi$ between the twin boundary and applied field is small, each vortex follows the twin boundary over some distance, then proceeds to the next boundary, and follows the boundary again, resulting in a kinked vortex.
Assuming $l \ll d$ and taking into account the interaction between the vortices, the critical angle $(\Delta\phi)^{\ast}$ where the vortices are released from the twin boundaries and get straight is given by
\begin{equation}
(\Delta\phi)^{\ast} \simeq \left[\frac{\pi\ln(\kappa\sqrt{\Gamma})}{2\sqrt{3}} \frac{2\Delta\epsilon_l}{\epsilon_l}\right]^{1/2} \left[\frac{l}{d}\right]^{3/2},
\end{equation}
where $\kappa$ and $\Gamma$ are the Ginzburg-Landau parameter and the mass anisotropy ratio ($m_c/m_{ab}$), respectively.
$\epsilon_l$ and $\Delta\epsilon_l$ are the line tension of a vortex and its reduction when it is trapped in the twin boundary, respectively.
Because $\Delta\epsilon_l \propto t(1-t)$ and $\epsilon_l \propto (1-t)$, where $t$ is a reduced temperature $T/T_c$, $\Delta\epsilon_l / \epsilon_l \propto T$.
Because $l \propto B^{-1/2}$, the temperature and field dependence of the critical angle is given by $(\Delta\phi)^{\ast} \propto T^{1/2}B^{-3/4}$.
Figure 6(b) confirms this relation, demonstrating the kinked-vortex scenario.

Finally, we mention a recent work \cite{Zhou23MTP}, where the magnetic torque in FeSe was measured as a function of the in-plane field angle in the mixed state.
The authors reported that the irreversible torque showed a peak when the field was parallel to the orthorhombic $a$ or $b$ axis, which corresponds to $\langle$110$\rangle_t$.
Although the peak suggests the enhanced pinning for this field direction, the reported direction differs from ours ($\langle$100$\rangle_t$) by 45$^{\circ}$.

\section{Conclusion.}
We have shown from bulk transport measurements that twin boundaries in FeSe pin vortices.
The width of the associated dip in $R_c(\phi)$ curves vary as $T^{1/2}B^{-3/4}$ as expected from the Blatter theory developed for YBCO \cite{Blatter91PRB}.
In the case of YBCO, the initial slope of the upper critical field $B_{c2}$ was so high that the temperature range where resistance dip measurements like the present ones could be performed was very limited.
In this study, by virtue of the relatively small initial slope of $B_{c2}$ in FeSe, we could demonstrate this relation for a meaningful temperature range.
The knowledge about the twin-boundary vortex pinning that we acquired in this study may be useful in improving superconducting wires and devices.
Further, it may be of use to future topological quantum computing.

\begin{acknowledgments}
This work was supported by Grant-in-Aid for Scientific Research on Innovative Areas ``Quantum Liquid Crystals'' (No. JP19H05824, JP22H04485), Grant-in-Aid for Scientific Research(A) (No. JP21H04443, JP22H00105, JP23H00089), Grant-in-Aid for Scientific Research(B) (No. JP22H01173), Grant-in-Aid for Scientific Research(C) (No. JP22K03537), and Fund for the Promotion of Joint International Research (No. JP22KK0036) from Japan Society for the Promotion of Science.
MANA is supported by World Premier International Research Center Initiative (WPI), MEXT, Japan.
TT acknowledges Haruhisa Kitano for valuable comments.
\end{acknowledgments}

\appendix
\section{Samples \#1 and \#2.}
The output voltage from sample \#1 was very noisy, probably because of bad electrical contacts and/or wires, and hence no meaningful data were obtained.
For sample \#2, $R_c(\phi, \theta)$ data were successfully recorded at $T$ = 8 K and $B$ = 2.2 T, and an $R_c(\phi)$ curve with dips, which is similar to that for sample \#3, was obtained (Fig. 7).
However, at higher magnetic fields, the out-of-phase component of the lock-in output increased largely, likely because of wire vibration, and hence no reliable data were obtained at higher magnetic fields and lower temperatures.

\begin{figure}
\includegraphics[width=8.6cm]{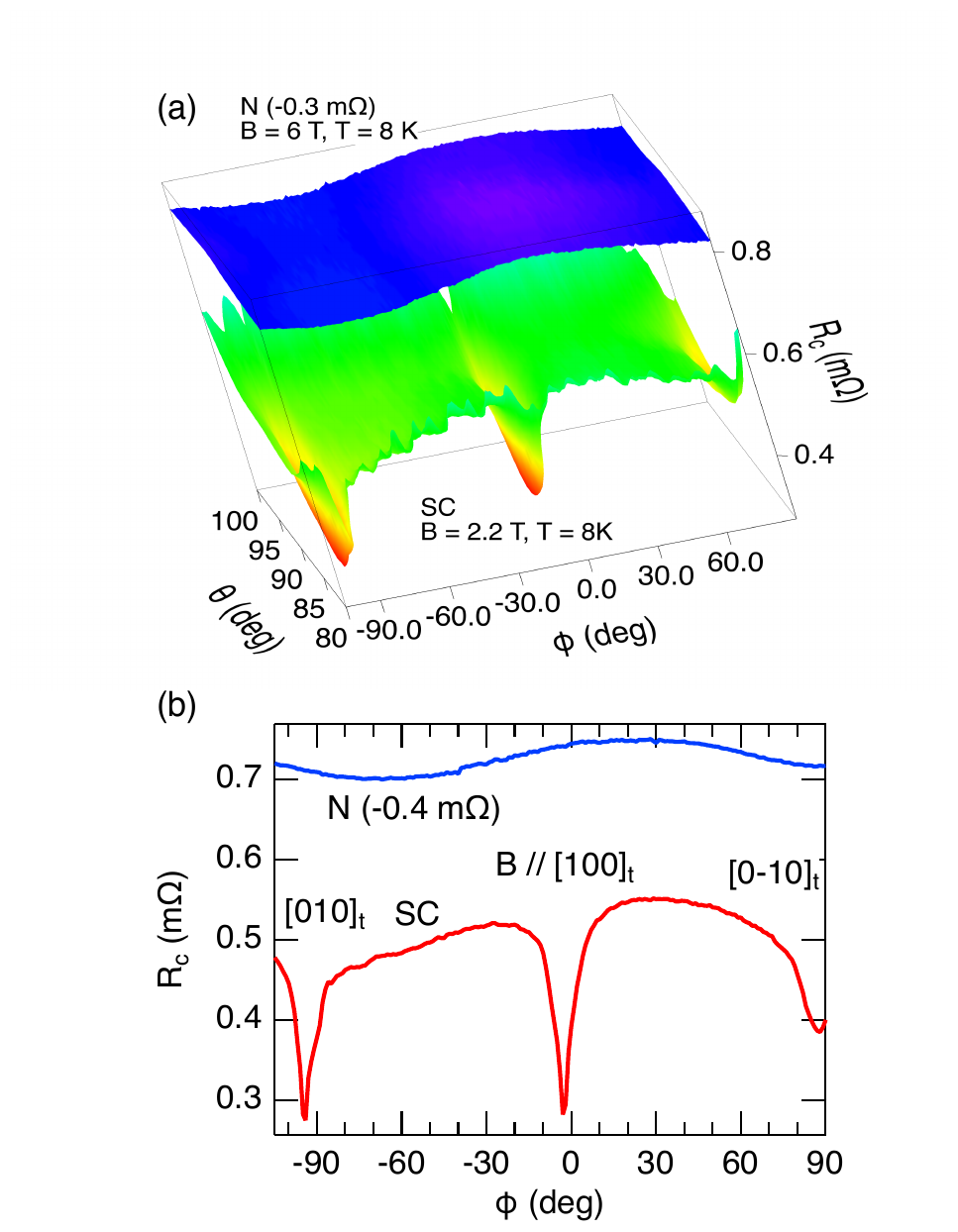}
\caption{\label{Sample2} Interlayer resistance $R_c$ of sample \#2 in magnetic fields.
(a) Resistance as a function of $\phi$ and $\theta$ measured at $B$ = 2.2 T and $T$ = 8 K in the superconducting transition region (SC) compared to that at $B$ = 6 T and $T$ = 8 K in the normal state (N).
The latter is offset by -0.3 m$\Omega$.
(b) Resistance for in-plane fields as a function of $\phi$ for the superconducting transition region (SC) and normal state (N).
The normal state curve is offset by -0.4 m$\Omega$.
}
\end{figure}

\begin{figure}
\includegraphics[width=8.6cm]{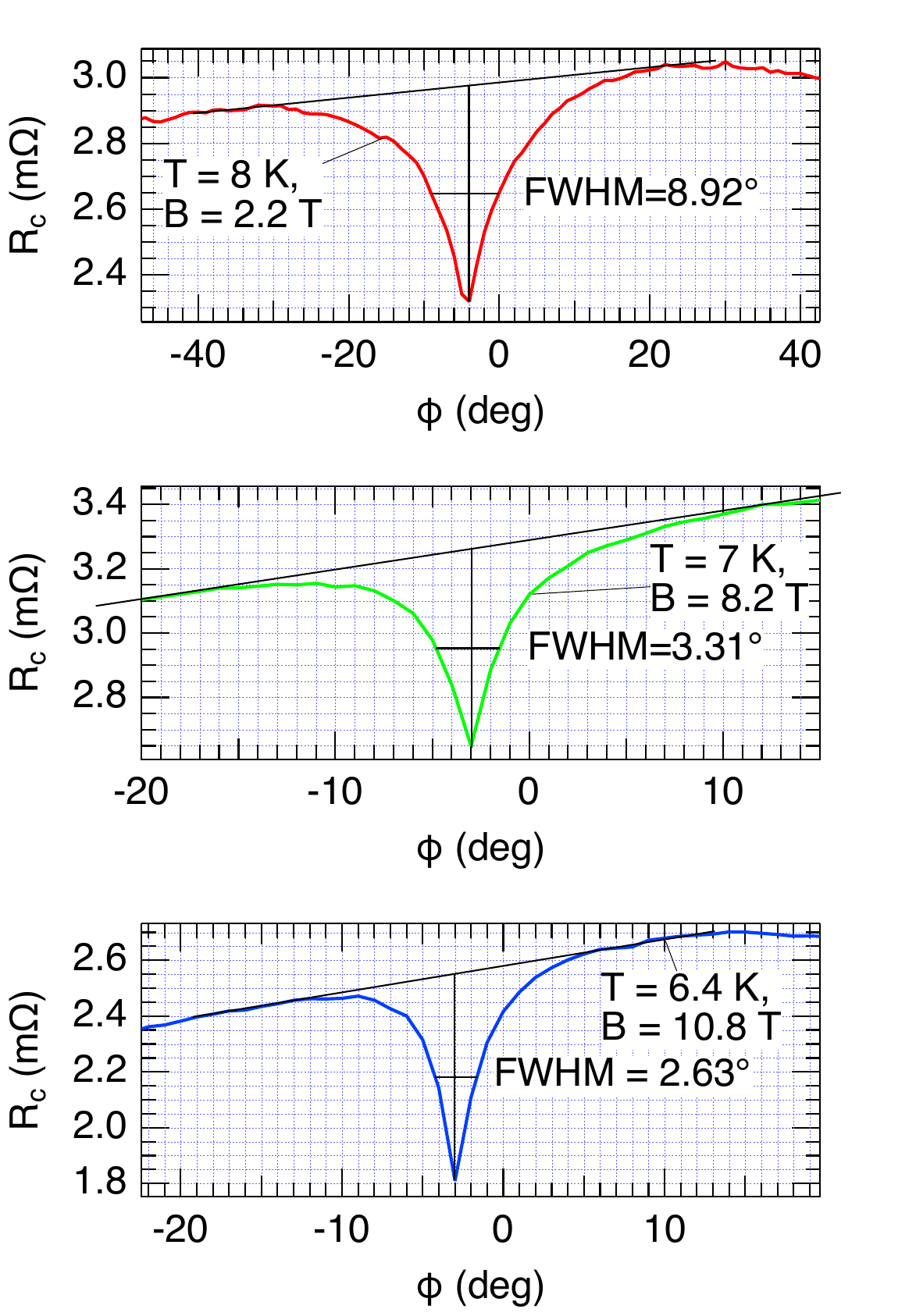}
\caption{\label{FWHM} The estimation of the FWHM of the $\phi \sim 0$ dip in sample \#3 is illustrated for (a) $T$ = 8 K and $B$ = 2.2 T, (b) $T$ = 7 K and $B$ = 8.2 T, and (c) $T$ = 6.4 K and $B$ = 10.8 T.
}
\end{figure}

\section{FWHM estimation.}
To estimate the FWHM of the $\phi \sim 0$ dip in sample \#3, we assumed a linear background as shown in Fig. 8.

\end{document}